\documentstyle[11pt]{article}

\relax
\catcode`\@=11

\newcommand{\e}{{\rm e}}                               
\newcommand{\ii}{{\rm i}}                               
\newcommand{\dd}{{\rm d}}                                
\newcommand{\half}{{1 \over 2}}                           
\newcommand{\Lie}{{\cal L}}
\newcommand{\RR}{{\cal R}}

\newcommand{\DD}{{\cal D}}
\newcommand{\doo}{{\partial}}
\newcommand{\sgn}{{\rm sign\  } }
\newcommand{\ind}{{\rm ind\ }}
\newcommand{\Ch}{{\rm Ch   }  }
\newcommand{\Td}{{\rm Td  }  }

\newcommand{\abs}[1]{{\vert {#1} \vert}}

\newcommand{\Tr}{{\rm Tr\ }}
\newcommand{\Det}{{\rm Det\ }}
\newcommand{\Pf}{{\rm Pf \  }}
\newcommand{\Str}{{\rm Str \  }}

\newcommand{\ra}{\rightarrow}

\newcommand{\ket}[1]{{\vert#1\rangle}}

\newcommand{\be}{\begin{equation}}
\newcommand{\ee}{\end{equation}}
\newcommand{\ben}{$$}
\newcommand{\een}{$$}
\newcommand{\ba}{\begin{eqnarray}}
\newcommand{\ea}{\end{eqnarray}}

\newcommand{\eg}{{e.g.}}

\begin{document}

\begin{titlepage}
\begin{flushright}
UU-ITP 14-1995 \\ hep-th/9508127
\end{flushright}

\vskip 0.5truecm

\begin{center}
{ \large \bf
ON LOCALIZATION AND REGULARIZATION
\\  }
\end{center}

\vskip 1.0cm

\begin{center}
{\bf Mauri Miettinen \\
\vskip 0.4cm
{\it Department of Theoretical Physics, Uppsala University
\\ P.O. Box 803, S-75108, Uppsala, Sweden \\}
\vskip 0.4cm}
\end{center}

\vskip 1.0cm

\rm
\noindent
Different regularizations are studied in localization of path
integrals. We discuss the effect of the choice of regularization by
evaluating the partition functions for the harmonic oscillator and the
Weyl character for SU(2). In particular, we solve the Weyl shift
problem that arises in path integral evaluation of the Weyl character
by using the Atiyah-Patodi-Singer $\eta$-invariant and the Borel-Weil
theory.

\vfill

\begin{flushleft}
\rule{5.1 in}{.007 in}\\
$^{*}${\small E-mail: $~~$ {\small \bf
mauri@rhea.teorfys.uu.se} $~~~~$
}
\end{flushleft}

\end{titlepage}

\section{Introduction}

Quantum localization is a generalization Duistermaat-Heckman theorem
\cite{dh} to infinite dimensions.  This theorem states that if the
Hamiltonian $H$ generates a global circle, or, more generally a torus
action in the phase space $\Gamma$ then the canonical partition
function is given exactly by the saddle-point approximation around the
critical points of $H$. Extensions to calculation of quantum
mechanical partition functions using phase space path integrals have
been represented in \eg \cite{us}.

We shall first consider basic ideas of localization. Then we shall
carefully regularize the pertinent functional determinants arising
from the path integrals. There is an ambiguity in choosing the
regularization scheme because of the spectral asymmetry of first order
differential operators. Therefore, the result depends on the
regularization as in the case of quantum mechanical anomalies.

Finally, we are going to apply our localization to the quantization of
the simple harmonic oscillator and to the evaluation of the Weyl
character of spin. We shall notice that different regularizations give
different energy spectra for the harmonic oscillator. We also show
that the continuum coherent state path integral yields directly the
correct character for spin if we choose an appropriate
regularization. In particular, we will consider the relation of
character formulae to the Borel-Weil theory which constructs the
irreducible representations of a Lie group as holomorphic
functions. Using this theory we relate the character formulae to the
equivariant index of the Dolbeault complex. The result is that the
path integral yields directly the correct character without an
explicit Weyl shift of the highest weight.
\\

\section{Localization of Phase Space Path Integrals}

We are interested in exact evaluation of phase space path integrals
(partition functions) of the form
\be\label{Zcanonical}
Z(T) = \int_{L \Gamma} {\DD}x \ \Pf \Vert \omega_{ab}(x) \Vert \exp \left( \ii
\int_0^T \dd t \left[ \vartheta_a {\dot{x}}^a - H(x) \right] \right)  \;.
\ee
where $\{x^a\} $ are local coordinates in $\Gamma$, $\Pf \Vert
 \omega_{ab} \Vert$ is the Liouville measure factor, $\vartheta_a$ the
 symplectic potential and $\omega_{ab} = \doo_a
\vartheta_b - \doo_b \vartheta_a$. The integration is performed over the loop
space $L\Gamma$ consisting of the phase space loops. The integrability
condition \cite{woodhouse} requires that $$\int_{\Sigma} \omega = 2
\pi n$$ for any 2-cycle $\Sigma$ in $\Gamma$ so that the path integral
is single valued. We introduce anticommuting variables $\psi^a$ to
write $\Pf
\Vert \omega_{ab} \Vert $ as a path integral
\be\label{ZSUSY}
 Z(T) = \int {\DD}x {\DD}\psi \ \exp \left( \ii \int_0^T \dd t
\left[\vartheta_a {\dot{x}}^a - H(x) + \half \psi^a
\omega_{ab} \psi^b \right]
\right) \; .
\ee
The boundary conditions are periodic also for the fermions, since they
are a realization of the differentials of the bosonic coordinates.

We interpret the path integral (\ref{ZSUSY}) in terms of equivariant
cohomology in $L\Gamma$. From the bosonic part of the action we get a
Hamiltonian vector field in $L\Gamma$
\ben
{\chi}_S^a = \dot{x}^a - \omega^{ab} \doo_b H
\een
whose zeroes define the Hamilton's equations.  The equivariant
exterior derivative in $L\Gamma$ is
\ben
\dd_S = \dd + \iota_S  \;,
\een
where $\iota_S$ denotes the contraction along the vector field
$\chi_S$.
The square of $\dd_S$ is the loop space Lie derivative
\ben%
\Lie_S = \dd \iota_S + \iota_S \dd  \sim {\dd \over \dd t} - \Lie_H \;.
\een
The action $S_{\rm B} + S_{\rm F}$ is supersymmetric under the
infinitesimal loop space supersymmetry transformations that are
parametrized by a gauge fermion $\delta \Psi$:
\ba\label{SUSY}
x^a \ra x^a + \delta \Psi \dd_S x^a & = & x^a + \delta \Psi
\psi^a , \cr \psi^a \ra \psi^a + \delta \Psi \dd_S \psi^a & = & \psi^a +
\delta \Psi \chi_S^a \;.
\ea
This implies that that the action is equivariantly closed: $$\dd_S
(S_{\rm B} + S_{\rm F}) =0 \;.$$ By an analogue of Fradkin-Vilkovisky
theorem \cite{bfv} one can show that  the path integral remains
intact if we modify the action by $S \ra S+\dd_S \Psi$ where $\Psi$
satisfies the Lie derivative condition
\be\label{EquivCohomology}
\dd_S^2  \Psi = {\Lie}_S \Psi =0 \;.
\ee
In the limit $\lambda \ra 0$ the path integral
\be\label{Z_lambda}
Z_{\lambda} (T) = {\cal} \int {\DD}x^a {\DD}\psi^a \
\exp \left( \ii \int_0^T \dd t \left[\vartheta_a {\dot{x}}^a - H(x^a)
+ \half \psi^a \omega_{ab} \psi^b + \lambda \dd_S \Psi \right] \right)
\ee
reduces to (\ref{ZSUSY}) and $\lambda \ra \infty$ gives localization.

To construct a gauge fermion $\Psi$ we need a metric $g$ in the phase
space. The loop space Lie derivative condition (\ref{EquivCohomology}) is
satisfied if the metric $g$ in $\Gamma$ is invariant under the
Hamiltonian action of $H$
\be\label{InvariantMetric}
{\Lie}_H g = 0 \;,
\ee
which means that $\chi_H$ is a Killing vector field. This is a very
restrictive condition for the Hamiltonian: it must generate a global
U(1)-action in $\Gamma$. We can choose any metric which satisfies the
condition (\ref{InvariantMetric}) and average it over the group
action.

We will consider the following selections for the gauge fermion:
\ben
\Psi_1 = {1 \over 2} g_{ab} \dot{x}^a \psi^b
\een
gives localization to the constant modes which are points of the
manifold,
\ben
\Psi_2 = {1 \over 2} g_{ab} \chi_H^a  \psi^b
\een
 to the zeroes of $\chi_H$ which we assume to be nondegenerate and
isolated and
\ben
\Psi_3 = {1 \over 2} g_{ab} \chi_S^a  \psi^b
\een
to the classical trajectories.  For simplicity we use subscripts 1,2,3
in the actions and partition functions corresponding to the gauge
fermions $\Psi_{1,2,3}$.  The actions become
\ba
S_1 & = & \int_0^T \dd t \ \left[ \left( \vartheta_a - {\lambda \over
2} g_{ab} {\chi}_H^b \right) \dot{x}^a -H + {\lambda \over 2} g_{ab}
\dot{x}^a \dot{x}^b  + {\lambda \over 2} \psi^a \left( g_{ab}
{\doo}_t + \dot{x}^c g_{bd} {\Gamma}_{ac}^d \right) \psi^b + {1 \over
2} \psi^a \omega_{ab} \psi^b \right] \;, \cr S_2 & = & \int_0^T \dd t
\ \left[ \vartheta_a \dot{x}^a - H + {\lambda \over 2} g_{ab}
{\chi}_H^a \left( \dot{x}^b - \chi_H^b \right) + {\lambda \over 2}
\psi^a \doo_{a }\left( g_{cb} \chi_H^c \right) \psi^b + {1 \over 2}
\psi^a \omega_{ab} \psi^b \right] \;, \cr S_3 &= & \int_0^T \dd t
\left[\vartheta_a \dot{x}^a - H +{\lambda
\over 2} g_{ab} \chi_S^a \chi_S^b + {\lambda \over 2} \psi^a \doo_a (g_{cb}
\chi_S^c) \psi^b + {1 \over 2} \psi^a \omega_{ab} \psi^b \right] \;.
\ea

To take the limit $\lambda \ra \infty$ in path integrals we make the
decomposition to constant modes $x_0^a, \psi_0^a$ and to non-constant
modes $x^a_t , \psi^a_t$ and scale the non-constant modes by $1/
\sqrt{\lambda} $
\ba\label{scaling}
x^a(t) & = &  x_0^a +  {1 \over {\sqrt{\lambda}}}x^a_t  ,  \cr
\psi^a(t) & =  &\psi^a_0 +  {1 \over {\sqrt{\lambda}}}  \psi^a_t .
\ea
The Jacobi determinant is unity.  An expansion to a quadratic order
around the constant modes and the limit $\lambda \ra \infty$ gives a
Gaussian path integral
\be\label{xdot-Gaussian}
Z_1 = \int \dd x_0^a \dd \psi_0^a \ \exp \left[ -\ii T \left( H-
\half \psi^a_0 \omega_{ab} \psi^b_0 \right) \right] Z_{\rm fl,1} (T)
\ee
where the fluctuation path integral $Z_{\rm fl}(T) $ is a product of
fermionic and bosonic parts:
\ba\label{x-Fluctutation}
Z_{\rm F ,1} &=& \int \prod_t \dd \psi^a_t \ \exp \left\{-{ \ii \over
2}\int_0^T \dd t \ \psi^a_t g_{ab} {\doo}_t \psi^b_t \right\} \; , \cr
Z_{\rm B,1} & = & \int \prod_t \dd x^a_t \   \exp \left\{ { \ii \over
2}  \int_0^T \dd t \ x^a_t \left[  \RR_{ab}    {\doo}_t -g_{ab} {\doo}_t^2
\right] x^b_t   \right\}  \; .
\ea
Here $$\RR_{ab} = R_{ab} + \tilde{\Omega}_{ab}$$ is the equivariant
curvature with $R_{ab} $ the Riemannian curvature 2-form and $$
\tilde{{\Omega}}_{ab} = \half \left[ {\nabla}_{b}(g_{ac} \chi_H^c) -
{\nabla}_{a}(g_{bc} \chi_H^c) \right] $$ the momentum map
\cite{arnold} corresponding to ${\chi}_H$, $\nabla$ being the
covariant derivative.  $H$, $\omega$, $g$ and ${\RR}$ are evaluated at
the constant modes. The path integral $Z_2$ is given by a sum over the
critical points $\{x_i \}$ of the Hamiltonian:
\be\label{FPICritical}
Z_2 = \sum_{ x_i } \ { {\exp[-\ii TH] } \over { \Pf \Vert \doo_a
\chi_H^b \Vert } } Z_{\rm fl,2}(T) \;.
\ee
$Z_{\rm fl,2}$ is also  a product of fermionic and bosonic parts:
\ba\label{chi-Fluctutation}
Z_{\rm F,2}(T) & = & \int \prod_t \dd \psi^a_t \ \exp \left\{ {\ii \over 2}
\int_0^T \dd t \ \psi^a_t \doo_a (g_{bc} \chi_H^c) \psi^b_t \right\} \;, \cr
Z_{\rm B,2}(T) & =& \int \prod_t \dd x ^a_t \ \exp \left\{ {\ii \over 2}
\int_0^T \dd t \  x^a_t \doo_a (g_{bc} \chi_H^c) (\delta^b_d \doo_t -
\doo_d \chi^b_H ) x^d_t \right\} \;.
\ea
Here $g$ and $\chi_H$ are again evaluated at the constant modes.
Finally, the path integral $Z_3$ reduces to a sum over the
$T$-periodic classical trajectories
\be \label{Classical}
Z_3 =   \sum_{x_{\rm {cl}}} { 1 \over \Pf \Vert \delta_b^a
 \doo_t - \doo_b   \chi_H^a
\Vert } \exp[\ii S_{{\rm cl }}] \;.
\ee
In practice, it is usually a highly non-trivial problem to find the
$T$-periodic classical trajectories of a dynamical system
\cite{ArnoldConj}.
\\

\section{Regularization of Fluctuation Path Integrals}

In the following all the path integrals and determinants are evaluated
over periodic configurations for both the bosonic and fermionic
degrees of freedom. The primes will denote that we exclude the
constant modes. In real polarization the fluctuation parts in
$Z_{1,2}$ become
\ba\label{xdot-Z_fl}
Z_{\rm fl ,1} & = & {1 \over \sqrt{\Det ' \Vert \delta_b^a \doo_t -
{\RR}_b^a \Vert }} \;,
\cr
 Z_{\rm fl,2} &= & {1 \over \sqrt{\Det ' \Vert \delta_b^a \doo_t -
\doo_b \chi_H^a \Vert}} \;.
\ea
It is quite important to notice that in the reduced determinants one
index is covariant and another contravariant.

 In K\"ahler polarization the fluctuations parts are, using the
additional symmetries of the metric and the Riemann curvature tensor
\cite{nakahara},
\ba
Z_{\rm fl,1} & = & {1 \over \Det' \Vert \delta_a^b \doo_t -\RR_a^b \Vert
}\;, \cr
Z_{\rm fl,2} & = & {1 \over \Det ' \Vert \delta_a^b \doo_t -
\doo_a \chi_H^b  \Vert  } \;.
\ea
These determinants are taken over the holomorphic indices. By this we
mean the following: The relevant matrices can be block diagonalized
\ben
A = {\rm diag} \ (A_1 ,A_2  ,... ,A_N)
\een
with blocks
\ben
A_k = \pmatrix{
a^+_k  & 0 \cr
0 &  a^-_k  \cr
} \equiv \pmatrix{
a_k & 0 \cr
0 & - a_k \cr
} \;.
\een
The symbols $a^+_k$ and $a^-_k$ denote the holomorphic and
antiholomorphic eigenvalues of $A$, and we consider only the
eigenvalues corresponding to the holomorphic indices to the
determinant.

We have to choose a regularization scheme for the determinants. A
standard method is to apply $\zeta$- and $\eta$-functions. The
$\zeta$-function regularization does not directly apply to first-order
operators because they have an infinite number of negative
eigenvalues. To take them into account we define the $\eta$-function
for the first-order operator $B$ by
\ben
\eta_B (s) = \sum_{b_{n} \neq 0}
{\sgn} (b_{n}) \abs{b_{n}}^{-s} + {\rm dim \ Ker \ } B = {1 \over
{\Gamma\left({s+1} \over 2 \right)}}
\int_0^{\infty} \dd t \ t^{(s-1)/2} \ \Tr [B \exp(-t   B^2)] \;.
\een
 Analytical continuation to $s=0$ gives the
Atiyah-Patodi-Singer $\eta$-invariant \cite{egh} of $B$ that measures
the spectral asymmetry of $B$ and specifies the phase of
$\Det(B)$. The absolute value $\abs{\Det (B) }$ is regularized using
the formula
\ben%
\abs{\Det (B)}
= +\sqrt{\Det (B^2)} = + \exp\left[ - \half
\zeta_{B^2} ' (0)\right] \;.
\een

In real polarization we have to evaluate the square root of a
determinant of the antisymmetric operator $B = \doo_t - A $ where $A$
is an antisymmetric matrix. In our case $A$ is $\Vert \RR_a^b \Vert$
or $\Vert \doo_a \chi^b_H \Vert$.  By determining the spectrum of $B$
and applying $\zeta$-function regularization we get, up to an
inessential numerical normalization, the result
\be\label{Agenus}
 {1 \over {\sqrt{\Det'
\left( \doo_t - A \right) }} }
= \prod_{n= 1}^N { \abs{a_n /2 \over {\sin (a_n T/2 )}}}= {1 \over
 T^N} \hat{A}(TA) \;,
\ee
where we have defined the function of the matrix $X$  $$\hat{A}(X) =
\prod_n {x_n/2 \over \sin(x_n/2)} \;$$ where $x_n$ are the skew-eigenvalues
of $X$. The result is non-negative since the negative and positive
skew-eigenvalues appear in pairs. Therefore there is no ambiguity with
the spectral asymmetry.

Now we consider the determinants in K\"ahler polarization. It is
sufficient to consider the determinant of a block. Earlier we noticed
that the fluctuation path integrals reduce to the determinant of the
operator $B = \ii \doo_t -a$. The functional Pfaffian in
(\ref{Classical}) is also similar to this determinant. To regularize
$$\Det'(B) = \prod_{n \neq 0} \left( {2\pi n \over T} - a
\right)$$ properly we have to take into account that $B$ has and
infinite number of negative eigenvalues. Thus there is a problem with
the spectral asymmetry.

Therefore, we have to choose a regularization prescription which has a
relation to quantum mechanical anomalies. In the regularization of the
determinants it is not possible to maintain all the symmetries that
are present in the classical theory.  For example, Elitzur
et al. \cite{elirab} considered the corresponding fermionic problem
with antiperiodic boundary conditions. They evaluated the quantum
mechanical partition function for a Dirac fermion in an external gauge
field $A(t)$ in $0+1$-dimensions
\be
Z(T) = \int \DD \bar{\psi} \DD \psi \exp\left[ \ii \int_0^T \dd t \
\bar{\psi} \left ( \ii \doo_t -a \right) \psi \right] = \Det ( \ii
\doo_t -a ) \;.
\ee
where, because of the gauge invariance of the action only the constant
mode $a$ of $A(t)$ contributes.  The classical action has both the
invariance under large gauge transformations
\ba
a &\ra&  a + n 2 \pi /T \cr
\psi  &\ra&  \psi
\ea
and  the charge conjugation invariance
\ba
a &\leftrightarrow&  -a \cr
\psi  &\leftrightarrow  &-\bar{\psi} \;.
\ea
However, when regularizing the determinant one has to choose which
symmetry one wants to maintain, which leads to a global anomaly.  Here
we have an analogous situation.  It is not {\em a priori} clear what
the result of the regularization should be, and there is a genuine
ambiguity.

Since the zeroes of the determinant are at $aT = 2\pi n $, the
determinant must be proportional to $$ {\sin( a T/ 2 )
\over a/2} \;.  $$  The proportionality factor can be any function
without zeroes that is, the exponent function.  The determinant is
therefore, up to an irrelevant constant,$$\Det(\ii \doo_t -a ) = {
\sin( a T/ 2) \over a/2} \exp\left(\ii \phi aT \right) $$
with a phase $\phi$ whose natural values turn out to be $0$ and $\pm
1/2$ since they yield the (anti)symmetries of the product under $a
\leftrightarrow -a $ and $a \ra a + 2\pi n /T$. However, there is
a minor subtlety: in our localization formulae the zero modes are
absent and this destroys these symmetries. Nevertheless, we may still
consider the residual symmetries. The choice $\phi=0$ corresponds to
neglecting the spectral asymmetry and choosing the (anti)symmetry $a
\ra -a$ to be unbroken. In this regularization scheme the inverse
determinant is simply
\ben%
{ 1 \over \Det'( \doo_t - A) } = {1 \over T^N} \hat{A}(T A)
\een
This is the result that usually appears in literature. However, there
is another possibility. The values $\phi = \pm 1/2$ correspond to
maintaining the symmetry $a \ra a + 2\pi n /T$ and taking into account
the spectral asymmetry by the Atiyah-Patodi-Singer
$\eta$-invariant. This yields \ba\label{ComplexDet} {1 \over {{\Det
'}(\doo_t - A)}} &=& \prod_{n= 1}^N {{a_n/2} \over {\sin \left( a_n T
/ 2 \right)}} \exp \left( \ii a_n T/2 \right) = {1 \over T^N} {\Td }(T
A)
\ea
where we have defined the following function of the matrix $X$
\ben
\Td (X) = \prod_n { x_n /2 \over \sin(x_n /2) } \e^{\ii x_n /2} \;.
\een
We take only the eigenvalues corresponding to the holomorphic indices
to the determinant.

Let us now write down the resulting localization formulae. The
localization to constant modes yields the expression
\be\label{ConstantModes}
Z_1 (T) = {1 \over T^N} \int  \dd x_0^a \dd \psi_0^a \
{\Ch}\left[ -\ii T(H- \omega)
\right]  \cases{
  \hat{A}({T \RR }) \cr \Td ({T \RR }) } \;.
\ee
We have defined equivariant generalizations \cite{bgv} of the
conventional characteristic classes known as equivariant $\hat{A}$-
and Todd-genus, and identified the exponential with the equivariant
Chern class.  When $H=0$ they reduce to the conventional
characteristic classes and the result is a topological invariant.  The
localization to the critical points $\{ x_i\}$ of the
Hamiltonian gives the result
\ba\label{CriticalPoints}
Z_2 (T) = {1 \over T^N} \sum_{x_i} {\exp(-\ii TH ) \over \Pf(\doo
\chi_H^{~} )}
\cases{
\hat{A}(T \doo \chi_H^{~}  ) \cr \Td (T \doo \chi_H^{~} )  } \;.
\ea
We must use local coordinates in the evaluation of the determinants
when  localizing to the critical points of the Hamiltonian.
Finally, the localization to $T$-periodic classical trajectories yields
\ba\label{WKB}
Z_3 (T) = {1 \over T^N} \sum_{x_{\rm cl}} {\exp(\ii S_{\rm cl} )
\cases{
\hat{A}(T \doo \chi_H^{~} ) \cr \Td (T \doo \chi_H^{~} )  }} \;.
\ea

\section{ Harmonic Oscillator}

Now we show that the localization formulae yield the correct partition
function for the harmonic oscillator in a flat phase space.  The path
integral for it is Gaussian and in principle there is no reason to
apply localization to it. However, it is reasonable to check by some
simple examples that our assumptions and derivations are valid.  In
particular, we will show that the choice of the metric in the phase
space is not relevant, contrary to claims in literature \cite{dlr}. It
is also illustrative to consider the significance of the
regularization schemes we have used.

 In real polarization the Hamiltonian is $H= \half \left( p^2 +q^2
\right)$ and the symplectic 2-form is $\dd q \wedge \dd p . $ The
coherent state representation (K\"ahler polarization) requires some
further investigation, since we have to fix an operator ordering
prescription. In terms of creation and annihilation operators the
normal and symmetric ordered Hamiltonians are, respectively,
\ba\label{Ordering}
H_{\rm n} & =& : {1 \over 2} \left( a^{\dagger}a + a a^{\dagger}
\right) : = a^{\dagger}a \;, \cr H_{\rm s} &=& a^{\dagger} a + \half
\; .
\ea
The symmetric ordered Hamiltonian has an explicit zero point energy
$E_0 = 1/2.$

To apply the localization formulae we must choose a metric in the
phase space and calculate the equivariant curvature and the
derivatives of the Hamiltonian vector field. If the Lie-derivative
condition $\Lie_H g = 0$ is satisfied we can start from an any smooth
metric in the phase space and average it. So we may choose a constant
metric $$g = \pmatrix{ 1 & 0 \cr 0 & 1 \cr }.$$ The non-zero
components of the equivariant curvature are
\ben
\RR^p_q = -  \RR^q_p = 1 \;.
\een
The localization formula (\ref{ConstantModes})
yields the result
\ba
Z(T) & = & \int \dd p \ \dd q \ \dd \psi^q \ \dd \psi^p \ \exp \left[
 -\ii T \left( {1 \over 2} p^2 + \half q^2 - \psi^p \psi^q \right)
 \right ] {1 \over 2\sin ( T/2)} \cr & \sim & {1 \over 2 \sin( T/2) }
 = \sum_{n=0}^{\infty} \exp [\ii (n + 1/2 )T ]
\ea
which is the correct partition function with the zero-point energy
$E_0 = 1 /2 $.

Let us now digress slightly to discuss the result. In \cite{dlr}
Dykstra, Lykken and Reiten analyzed this problem and they noticed a
dependence on the metric. What they did not notice was that the index
structure of the equivariant curvature is $\RR_a^b$ and therefore it
is invariant under global scalings of the metric. Furthermore, they
used a metric which in polar coordinates near the origin behaves like
\ben
\dd s^2 = \dd r^2 + c r^2 \dd \phi^2 \;.
\een
This is a metric on a cone, not on a plane when $c \neq 1$ and is not
smooth, nor even continuous at the origin. Therefore it is not
surprising that their energy levels depend on the parameter $c$ which
represents the tip angle of the cone. From this we indeed see that we
cannot choose an arbitrary invariant metric, since it has to respect
the topology of the phase space.

The localization to the critical points of the Hamiltonian
(\ref{CriticalPoints}) yields also the correct result.  The only zero
of $\chi_H^{~} $ is the origin of the phase space, which gives
\ben
Z(T) = {1 \over \Pf \pmatrix{ 0 & 1 \cr -1 & 0 \cr }} { {1 /2}
\over {\sin ( T/2)}} = {1 \over { 2 \sin(T/2)} } \;.
\een
If we want to apply the localization the classical trajectories we
must classify all the $T$-periodic classical trajectories.  If $T \neq
2 \pi n$ the problem reduces to the localization to the critical
points of the Hamiltonian. However, if $T = 2 \pi n $ the zeroes of
$\chi_S$ are not isolated and we have to use a degenerate version of
the localization formula to the classical trajectories \cite{Palo}

We now consider the harmonic oscillator in the K\"ahler
polarization. We will only discuss the localization formulae to
constant modes. The reasoning is similar with other formulae.  There
are four cases to consider: the localizations with the $\hat{A}$-genus
and Todd-genus using two different orderings. We will only list the
spectra we obtain. The use of $\hat{A}$-genus yields the spectra $E_n
= n+1/2$ (normal ordering) and $E_n = n+1$ (symmetric ordering). The
Todd-genus gives the results $E_n = n$ (normal ordering) and $E_n =
n+1/2$ (symmetric ordering). The first and fourth results have the
correct zero-point energy.  From this example we see that to get
correct results from the path integral we do need some additional
information other than the classical action and boundary conditions:
we must choose a regularization scheme that gives physically correct
results.

\section{ Character for SU(2)}
We shall now  use our localization formulae to derive the
Kirillov and Weyl character formulae for Lie groups \cite{bgv}. The
character formula for SU(2) has been widely discussed in literature
\cite{afs, stone, nielsen}. However, there has been some
controversy about the Weyl shift problem: the path integral usually
gives almost the correct character up to the substitution
$j \ra j+1/2$. We show that the coherent state path integral and the
localization formulae with the Todd-genus directly yield the correct
character. In this calculation we use the continuum version of the
coherent state path integral and show that this also yields the
correct result, contrary to discussions in literature
\cite{funahashi}.

To motivate the use of Todd-genus we relate the character of a simple
Lie group $G$ in the highest weight representation $\lambda$ to the
index of the twisted Dolbeault complex on the coadjoint orbit $O_f$
\cite{perelomov} of the group. The Borel-Weil theory  \cite{ stone, alsiwi}
constructs the irreducible representations of $G$ as holomorphic
sections of a line bundle $L$ that is associated to a principal bundle
$ G \ra G/T_G \sim O_f$ where $T_G$ is the Cartan torus of $G$
\cite{alsiwi}. The holomorphic sections of this line bundle
 (coherent states) form the basis for the irreducible
representation. The connection 1-form on $L$ is the symplectic
potential
\ben%
\vartheta = {\doo{F} \over \doo z^k } \dd z^k - {\doo{F} \over \doo
\bar{z}^k} \dd \bar{z}^k \;
\een
where $F$ is the K\"ahler potential on $O_f$.  It can be shown that
the twisted Dolbeault operator $\bar{\doo}_L = \bar{\doo} +
\vartheta_{\bar{z}}$ annihilates the normalized coherent states
$\ket{z}$ and therefore $\ket{z} \in H^{0,0} (O_f ,
L)$. If we can prove that all the other cohomology
groups are trivial, \eg by Lichnerowicz vanishing theorem
\cite{bgv}, we conclude that the dimension of the highest weight
representation $R_{\lambda}$ is $\dim H^{0,0} (O_f, L)$.
Consequently, this is equal to the index of the twisted Dolbeault
complex. The Riemann-Roch-Hirzebruch index theorem relates this
analytical index to the topological invariant
\be\label{TopDolbeault}
{\ind} \ \bar{\doo}_L = {\dim}\ R_{\lambda} = \int_{O_f}
{\rm Td}(O_f) \wedge {\Ch}(L) \;.
\ee
Indeed, we notice that the localization formula (\ref{ConstantModes})
with $H=0$ represents this index provided we use the Todd-class. For
SU(2) we obtain the known result for the dimension of the
spin-$j$- representation
\ben
\dim R_j =  \ind \ \bar{\doo}_L =   2j +1 \;,
\een
This is the correct result  without the explicit Weyl shift by the
Weyl vector $\rho = 1/2$.

We shall now use an equivariant version of the index theorem to derive
the character formulae.  The character of an element in the Cartan
subalgebra is the partition function for the Hamiltonian $H$ that
represents it on $O_f$:
\be\label{character}
\chi (\beta) = \Str \exp[- \ii  T H ]   \;.
\ee
To make a relation to the Dolbeault index we write this as an
equivariant index (character index, G-index, Lefschetz number)
\cite{bgv}. One can show that the Laplacians $\bar{\doo}_L^{\dagger}
\bar{\doo}_L$ and $\bar{\doo}_L \bar{\doo}_L^{\dagger}$ have
equal non-zero eigenvalues. If all other comohomology classes except
 $H^{0,0}$ are trivial, as we presume, $\bar{\doo}_L^{\dagger} $ does
 not have zero modes. Consequently, we can write the trace as an
 equivariant index:
\ba
\ind_H (\bar{\doo}_L, T) &\equiv& \lim_{\beta \ra \infty} \Tr
\e^{- \ii T H} (\e^{- \beta \bar{\doo}_L^{\dagger} \bar{\doo}_L} -
\e^{- \beta \bar{\doo}_L \bar{\doo}_L^{\dagger}} )\cr
 & =& \lim_{\beta \ra \infty}
\Str \exp[- \ii T H] \exp \left[- \beta \pmatrix{ \bar{\doo}_L^{\dagger}
\bar{\doo}_L & 0 \cr  0 & \bar{\doo}_L \bar{\doo}_L^{\dagger} }  \right]
\ea
Only the zero modes contribute to the trace. The expression is also
independent of $\beta$. Therefore, in the limit $\beta \ra 0$, all we
are left with are the zero modes of $\bar{\doo}_L$. Consequently, the
equivariant index is equal to the character
\ben
\ind_H (\bar{\doo}_L, T) = \Str \exp[- \ii T H]
\een
Thus, the character is the equivariant index of the twisted Dolbeault
complex and therefore we choose the localization with the Todd-class.

To derive the character formulae we apply standard methods to write
 $\Str \exp[ - \ii T H]$ as a coherent state path integral of the form
 (\ref{ZSUSY}).  Since we can choose an invariant metric on a
 coadjoint orbit \cite{perelomov} we can localize the path integral to
 classical trajectories, to constant modes or to critical points of
 the Hamiltonian. The two latter cases yield the Kirillov character
 formula \cite{kirillov} ($2N$ is the dimension of the orbit)
\be\label{Kirillov}
\chi (T) =  {1 \over T^N} \int_{O_f} {\Ch}\left[ -\ii  T (H - \omega)
\right]   \Td ({ T \RR^+}) \;,
\ee
and the Weyl character formula
\be\label{Weyl}
\chi (T) =  {1 \over T^N}  \sum_{z_i} {\exp(-\ii  T H ) \over
 \det^+ ( \doo \chi_H^{~}  ) } \Td (T \doo \chi_H^{~} ) \;,
\ee
respectively.  In (\ref{Weyl}) we have identified the Pfaffian in the
real polarization with the determinant over the holomorphic
eigenvalues of $\Vert \doo_a \chi^b_H \Vert $ and the summation is
over the critical points of the Hamiltonian or equivalently the Weyl
group.

As the only example we evaluate the character for SU(2). We write the
character as a coherent state path integral over the coadjoint orbit
$\rm{SU(2)/U(1)} \sim S^2$. We choose complex coordinates by
introducing the stereographic projection from the south pole. The
K\"ahler potential on the orbit with radius $j$ is $F = j
\log(1 +z \bar{z})$ from which we obtain the metric and the symplectic 1-
and 2-forms in the standard fashion. The integrability condition
requires $j$ to be a multiple of $1/2$: this is the topological
quantization of spin. The canonical realization for $H = J_3$ is
\ben
J_3 = -j {{1 -z\bar{z}} \over {1+z\bar{z}}}  \;
\een
and the  path integral for the character becomes
(\ref{ZSUSY})
\be\label{SU2Character}
\chi_j (T) =  \int \DD z \DD \bar{z} \DD \psi \DD \bar{\psi} \
\exp\left[ \ii j \int_0^T \dd t \left( \ii {\dot{z} \bar{z} -
z\dot{\bar{z}} \over 1 + z\bar{z}} + {1 -z\bar{z} \over 1 + z \bar{z}}
+ {2\ii \psi \bar{\psi} \over (1 +z\bar{z})^2} \right) \right]
\ee
with periodic boundary conditions. The Lie-derivative condition
(\ref{InvariantMetric}) is satisfied for $H =J_3$.  This path integral
is given exactly by  the WKB-approximation \cite{us,funahashi}.
The relevant quantities in the Kirillov formula (\ref{Kirillov}) are
\ba\label{PSInvariance}
H- \omega & = & j { 1 -z \bar{z} - \psi \bar{\psi} \over 1 + z \bar{z}
+ \psi \bar{\psi}} \;, \cr
\RR^+ &=& R^+ + \Omega^+ = {1 -z \bar{z} - \psi \bar{\psi} \over 1 +z \bar{z}
+ \psi \bar{\psi}} \;.
\ea
Now one can use the Parisi-Sourlas integration formula $$\int \dd z \ \dd
\bar{z} \ \dd \psi \ \dd \bar{\psi} F(z \bar{z} + \psi \bar{\psi}) = \pi
[F(\infty) - F(0)]  $$ which gives
\be\label{char}
\chi_j (T) = {\sin (j+1/2)T \over \sin (T/2)} = \sum_{m = -j}^j
\exp[ \ii m T] \:.
\ee
This is exactly the correct result without an explicit Weyl
shift. Also the Weyl formula (\ref{Weyl}) gives the correct result
when we use local coordinate charts in the vicinity of the critical
points. To get the correct north pole contribution we invert the
coordinates $z \ra 1/z, \bar{z} \ra 1/\bar{z}$. This also yields the
correct character (\ref{char}):
\ben
\chi_j (T) = { \exp[-\ii j T  ] \over 2 \sin(T/2) } \exp [- \ii T/2 ]
+ { \exp [-\ii j T (-1) ] \over 2 \sin(-T/2) } \exp[\ii  T /2 ] =
{\sin(j+1/2)T \over \sin(T/2) } = \sum_{m= -j}^j \exp[\ii m T ] \;.
\een
On the other hand, using $\hat{A}$-genus we obtain the result
\ben
\chi_j (T) = {\sin (j T) \over \sin(T/2) }
\een
which is the correct result up to the Weyl shift $j \ra j+1/2$.  So we
see that in the character formulae we have to use the Todd-genus
instead of $\hat{A}$-genus to directly get the correct result.
\\

\section{Conclusions}

We have considered phase space path integrals with the property that
the Hamiltonian generates an isometry of the phase space. Using
equivariant cohomology in the loop space we were able to reduce the
path integrals to finite dimensional integrals and sums.  We also
noticed that the results were not uniquely defined because of spectral
asymmetry. The choice of regularization yielded equivariant
$\hat{A}$- and Todd-classes.

We applied localization to the harmonic oscillator and to the
quantization of coadjoint orbits. We showed that localization produces
correct results for these systems. In addition, we derived Kirillov
and Weyl character formulae that produce correct characters for Lie
groups without the Weyl shift. We demonstrated this explicitly by
evaluating the character for SU(2). The explanation for the Weyl shift
was the same as in the case of the Coxeter shift \cite{bbrt} in
Chern-Simons theory, the $\eta$-invariant.

It would be interesting to apply our formalism to more complicated
systems such as loop groups and field theories. Also, it seems
possible  to use localization and equivariant cohomology to study
quantum integrability, generic supersymmetric theories and problems
in classical mechanics, as well. \\

{\bf Acknowledgements}
\\
We thank Prof. Antti Niemi for initiating this project and for
valuable discussions. We also thank A. Alekseev, A. Hietam\"aki and
O. Tirkkonen for discussions and comments.


\end{document}